\documentclass[letterpaper]{article} 
\usepackage{aaai2026}  
\usepackage{times}  
\usepackage{helvet}  
\usepackage{courier}  
\usepackage[hyphens]{url}  
\usepackage{graphicx} 
\usepackage{natbib}  
\usepackage{caption} 
\usepackage{algorithm}
\usepackage{algorithmic}

\usepackage{amsmath,booktabs,multirow,amssymb}

\usepackage{xcolor}
\newcommand{\answerYes}[1]{\textcolor{blue}{#1}} 
\newcommand{\answerNo}[1]{\textcolor{teal}{#1}} 
\newcommand{\answerNA}[1]{\textcolor{gray}{#1}}

\usepackage{newfloat}
\usepackage{listings}
\DeclareCaptionStyle{ruled}{labelfont=normalfont,labelsep=colon,strut=off} 
\floatstyle{ruled}
\newfloat{listing}{tb}{lst}{}
\floatname{listing}{Listing}
\title{GASTON: Graph-Aware Social Transformer for Online Networks}
\author{
    Olha Wloch,
    Liam Hebert,
    Robin Cohen,
    Lukasz Golab
}
\affiliations{
    University of Waterloo\\
    Waterloo, Ontario, Canada

}

\begin{document}

\maketitle

\begin{abstract}
Online communities have become essential places for socialization and support, yet they also possess toxicity, echo chambers, and misinformation.  Detecting this harmful content is difficult because the meaning of an online interaction stems from both what is written (textual content) and where it is posted (social norms). We propose GASTON (Graph-Aware Social Transformer for Online Networks), which learns text and user embeddings that are grounded in their local norms, providing the necessary context for downstream tasks. The heart of our solution is a contrastive initialization strategy that pretrains community embeddings based on user membership patterns, capturing a community's user base before processing any text. This allows GASTON to distinguish between communities (e.g., a support group vs. a hate group) based on who interacts there, even if they share similar vocabulary. Experiments on tasks such as stress detection, toxicity scoring, and norm violation demonstrate that the embeddings produced by GASTON outperform state-of-the-art baselines. 
\end{abstract}

 \begin{links}
 \link{Code}{https://github.com/olhawloch/gaston}
 \end{links}

\section{Introduction}

Online communities have become essential spaces for socialization and support, fostering connections for everything from niche hobbies to sensitive health discussions. However, pseudo-anonymity and the lack of physical cues can facilitate the spread of misinformation, toxicity, and the formation of echo chambers. Mitigating these harms would improve user well-being but is difficult because the meaning of an online interaction is shaped by the interplay of what is written (textual content) and where it is posted (social norms). 

\begin{figure}
  \centering
  \includegraphics[width=\linewidth]{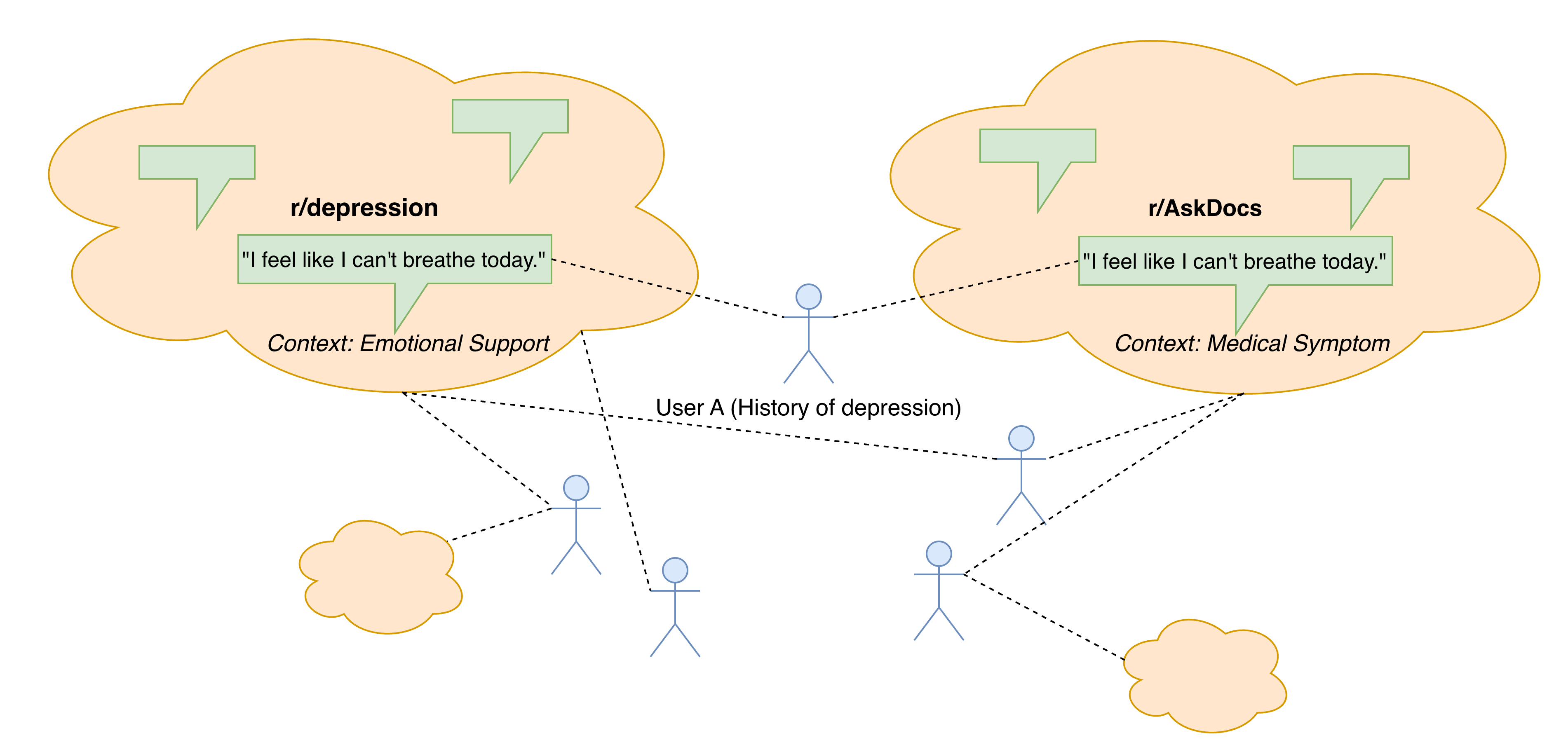}
  \caption{A fragment of the Reddit social platform}
  \label{fig:reddit_dynamics}
\end{figure}

For example, Figure~\ref{fig:reddit_dynamics} illustrates a fragment of the Reddit platform, with each community (subreddit) name indicating its topic; e.g., \texttt{r/AskDocs} is a community where doctors answer questions.  Here, User A participates in two communities. The meaning of `I feel like I can't breathe' varies by community context, serving as a cry for emotional support in \texttt{r/depression} versus a physical symptom description in \texttt{r/AskDocs}. A model that ignores this community signal fails to capture the post's intent.

Current approaches often treat different signals in isolation: text-only models analyze content but miss the social norms that define acceptable behaviour, while structure-only models map relationships but ignore the semantic content of discussions. Hybrid approaches attempt to bridge this gap but overlook local characteristics such as community norms.

To address this limitation, we propose GASTON (Graph-Aware Social Transformer for Online Networks), a graph learning framework that models connections between users, communities, and text. This enables grounding user and text embeddings in their local norms, providing the necessary context to accurately classify behaviour in downstream tasks.  The technical challenge addressed by GASTON is how to learn this context.  The core of our solution to this problem is a contrastive initialization strategy that pretrains community embeddings from user membership patterns, capturing the unique signature of a community's user base before the model processes any text. This allows GASTON to distinguish between communities (e.g., a support group vs. a hate group) based on who interacts there, even if they share similar vocabulary. Our contributions are as follows.

\begin{enumerate}

\item On the conceptual side, we propose to establish the community as a foundational contextual entity when modelling online interactions. 

\item On the implementation side, we embody the above concept in GASTON, a graph learning architecture that models a community's identity as a distinct learnable entity rather than a derivative of its textual content. This is aided by our novel contrastive initialization strategy. By learning structural representations for communities and fusing them with user and text embeddings, we provide the necessary context to ground online discourse.  

\item On the evaluation side, we test GASTON on socially-aware downstream tasks, including mental health stress detection, toxicity scoring, and norm violation detection. Our experiments demonstrate that the context-aware embeddings produced by GASTON outperform state-of-the-art baselines, particularly in tasks where social context is critical for classification, such as detecting norm violations. 

\end{enumerate}

\section{Related Work}

We categorize online interaction models by the signals considered: network structure, text, or both.  

Structure-only methods include node2vec \cite{grover2016node2vecscalablefeaturelearning} and community2vec \cite{martin2017community2vec}. By analogy to word2vec, the latter treats communities as `words' and users as `contexts'. For example, a user who comments in \texttt{r/nba} and \texttt{r/lakers} creates a co-occurrence pair that links these two communities. By building a co-occurrence matrix from all user participation on Reddit, the model learns an embedding vector for each community.  These structure-only representations capture semantic relationships. For instance, the embedding for \texttt{r/berlin} minus \texttt{r/germany} plus \texttt{r/unitedkingdom} is the closest embedding for \texttt{r/london}. This work establishes that user co-membership data is a rich source of information for modelling community relationships.

Text-only methods often fine-tune large language models such as BERT \cite{devlin2019bert} for a given social media task (e.g., HateBERT for hate speech detection \cite{caselli2021hatebert}). However, text-only hate speech detectors have been shown to misclassify reclaimed slurs used supportively within a specific community because it lacks the social norm context to interpret the word correctly \cite{hebert2023qualitative}.

Recently, One-Model-Connects-All (OMCA) incorporated both text and network structure \cite{ma2023one}.  OMCA uses a heterogeneous graph comprising three node types: communities, users, and text. These nodes are connected by edges representing their relationships, such as Text-posted-by-User and Text-belongs-to-Community. To initialize this graph, text nodes are encoded with a Language Model (XLNet) \cite{yang2019xlnet}, while user and community nodes are initialized by averaging the text embeddings they are connected to. A graph transformer is then pretrained to produce final embeddings by propagating information among the nodes.  GASTON builds upon this architecture by replacing averaging with a contrastive learning objective to better capture community identities and incorporate user-membership structural information.

The models described above serve as the foundation for a wide range of applications, from micro-level tasks such as hate speech classification to macro-level analyses of entire platforms, including political polarization \cite{waller2021quantifying}.

Notably, the strategy of fusing network structure with semantic content has been used in other complex domains such as healthcare. For example, Wu et al. (\citeyear{wu2021leveraging}) demonstrated the utility of incorporating structural knowledge into representation learning. They proposed a graph-based Hierarchical Medical Entity (HME) embedding framework that leverages the topology of medical knowledge graphs (e.g., the hierarchical relations between diagnoses and treatments) and textual descriptions. 

\section{Our Solution}

\subsection{Excecutive Summary}

\begin{figure}
  \centering
  \includegraphics[width=\linewidth]{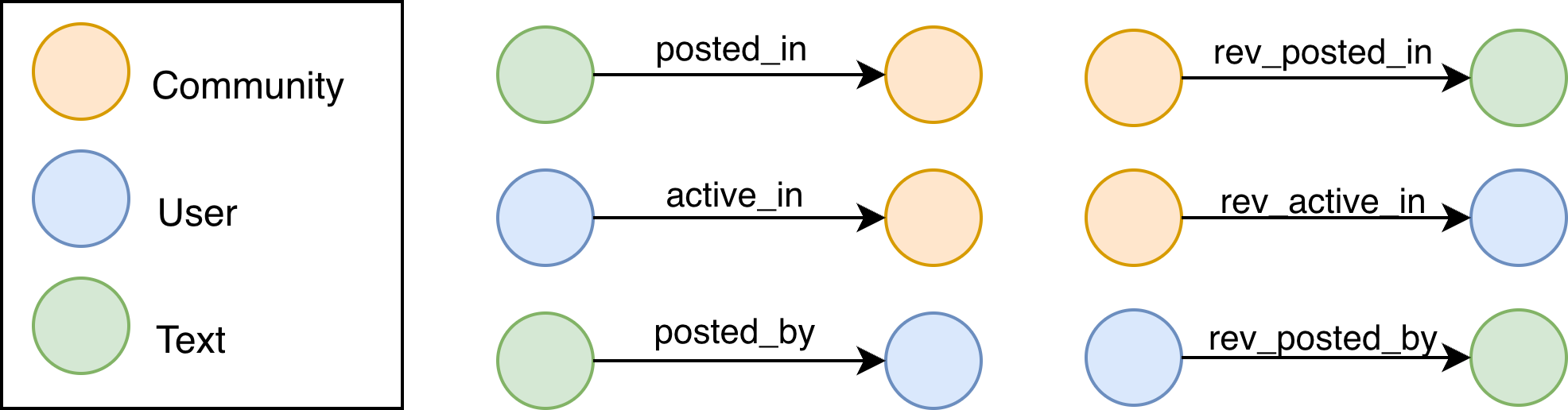}
  \caption{Node and edge types used in GASTON}
  \label{fig:GASTON_legend}
\end{figure}

GASTON uses a multi-layer Heterogeneous Graph Transformer (HGT) \cite{hu2020heterogeneous}, which operates on a heterogeneous graph $G = (V, E)$ containing multiple types of nodes $V$ and edges $E$.  As summarized in Figure~\ref{fig:GASTON_legend}, we define three types of nodes (users, text, and communities) and six types of edges (Text-PostedIn-Community, User-ActiveIn-Community, Text-PostedBy-User, and their reverse counterparts). Given a pretraining dataset (in our case, Reddit), we convert it to a heterogeneous graph as follows. We create a text node for each post and comment, a user node for each unique user, and a community node for each unique community. We then create edges linking text, user and community nodes, with a user connected to a community if they post or comment in it at least once. 

In general, graph neural networks are composed of \emph{message-passing} layers, in which each node aggregates the embeddings of its neighbours and uses this information to update its own embedding. By stacking these layers, a node's representation becomes progressively more aware of its wider structural context. This allows the network to learn a representation for each node that captures its structural context within the graph. HGTs go a step further and learn distinct aggregation and update functions for different edges, as is the case in our problem.

Before being processed by the HGT message-passing layers, each node $v$ must be associated with an initial embedding $X_v$.  For text nodes, we use the EmbeddingGemma pretrained embeddings \cite{google2024embeddinggemma} (specifically, we use the model's \texttt{retrieval-document} prompt to generate retrieval-optimized embeddings). For community and user nodes, instead of initializing their embeddings as the average embeddings of their posts, as in OMCA, GASTON treats these embeddings as learnable parameters. The message-passing layers of HGT then produce the final embeddings during a pretraining stage that reconstructs masked nodes and edges. 

\begin{figure*}
  \centering
  \includegraphics[width=0.68\textwidth]{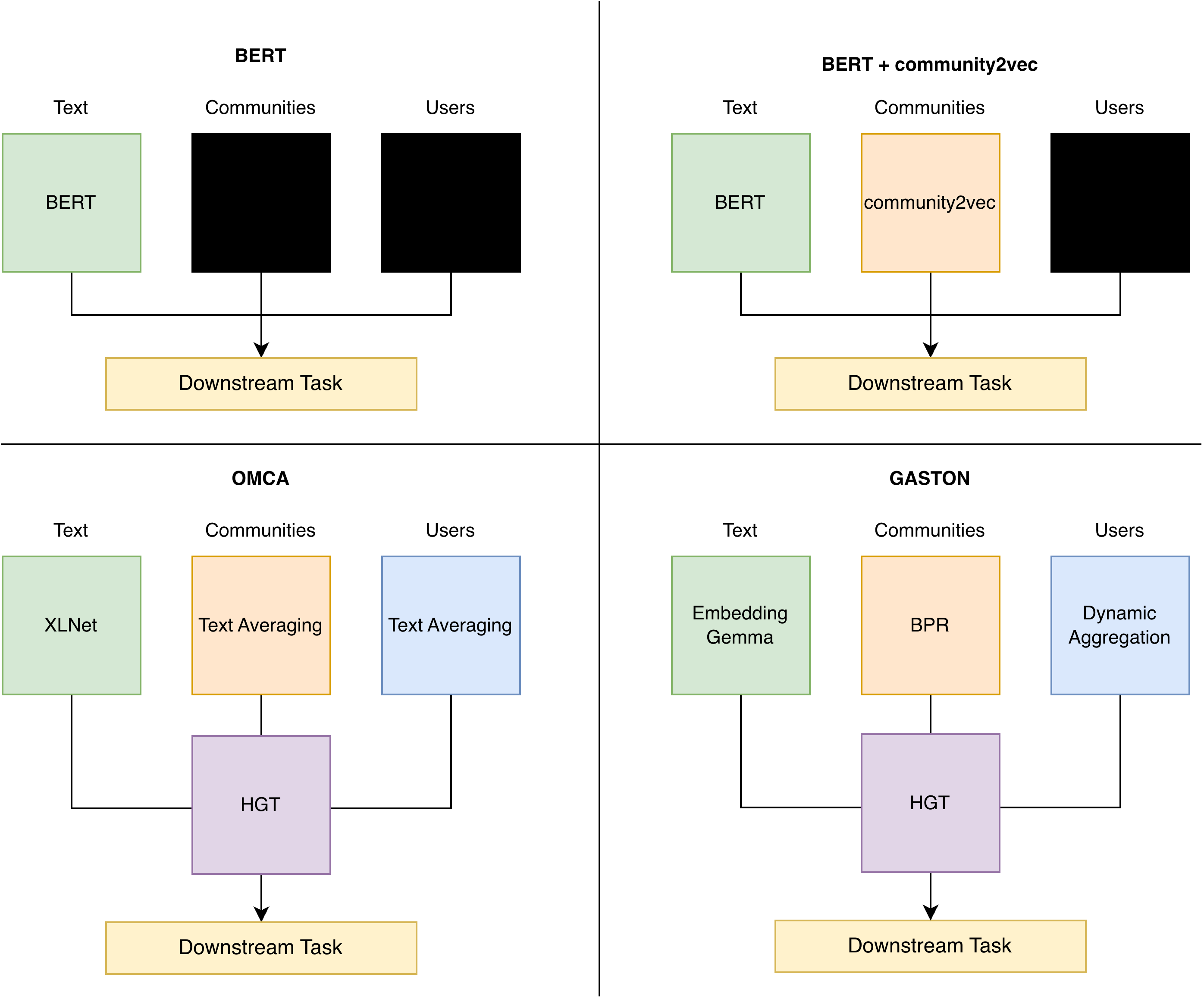}
  \caption{An architectural comparison of GASTON with prior work}
  \label{fig:model_comparison}
\end{figure*}

Figure~\ref{fig:model_comparison} compares GASTON to existing solutions. In the top-left corner is a text-only model that uses BERT embeddings as features for downstream tasks. A simple extension to add structure is to concatenate the corresponding community embedding (e.g., computed using community2vec) with the text embedding, as shown in the top-right panel. The state-of-the-art text-plus-structure model, OMCA, is shown on the bottom left, with its initial community and user embeddings corresponding to the average embeddings of the text posted within them and the text they authored, respectively, followed by graph transformer layers.  GASTON is illustrated in the bottom-right panel, showing two key advancements: the use of EmbeddingGemma, a language model optimized for generating embeddings, and the decoupling of the initialization of social nodes (users and communities) from that of text nodes via contrastive initialization.  In the remainder of this section, we describe the details of Bayesian Personalized Ranking (BPR) contrastive initialization for communities,  dynamic user aggregation, and the pretraining pipeline.

\subsection{Initializing Community Node Embeddings}
\label{sec:comm_node_embeddings}

We define a bipartite user-community interaction subgraph consisting of two disjoint sets of nodes,  users and communities, where edges only exist between sets. Inspired by collaborative filtering \cite{he2017neuralcollaborativefiltering}, we formulate a community recommendation problem. We first initialize trainable embedding matrices for all users and all communities. The model is then trained independently using a contrastive learning objective based on Bayesian Personalized Ranking (BPR) \cite{rendle2009bpr}. The process is as follows:

\begin{enumerate}
    \item A positive sample is created by selecting a (user, community) pair that exists in our interaction graph.
    \item A negative sample is formed by pairing the same user with a community they have not interacted with.
    \item The model computes a similarity score for both the positive and negative pairs.
    \item A ranking loss is used to optimize embeddings. This loss function pushes the similarity score of the positive pair to be greater than that of the negative pair.
\end{enumerate}

This process yields initial community node embeddings that are semantically organized based on user co-membership patterns, capturing structural relationships independently of any text content.

\subsection{Initializing User Node Embeddings}
\label{sec:dynamic_user_aggregation}

From the learned community embeddings, user node embeddings are then dynamically generated in each training batch. For each user in a given batch, we identify all communities in which they are active using a precomputed aggregation map. The user embedding is then computed as the average of the embeddings of their associated communities. To ensure efficiency during training, the mapping of users to their communities is precomputed and stored in an adjacency list before the training loop begins.

\subsection{Node Type Projections}

To process the learnable community embeddings alongside text embeddings, we implement a \emph{node-type tagging} layer. Before the first HGT message-passing layer, every node's input feature vector $X_{v}$ is projected into a shared hidden dimension $d$ via a type-specific linear transformation. For a node $v$ of type $\tau(v)$, its projection $H^{(0)}_{v}$ is calculated as:

\begin{equation}
    H^{(0)}_{v} = W_{\tau(v)} \cdot X_{v} + b_{\tau(v)}
\end{equation}
where $H_{v}^{(0)} \in \mathbb{R}^{d}$ is the initial projected hidden state for node $v$; $X_{v}$ is the initial embedding for node $v$ (derived from EmbeddingGemma for text, or BPR initialization for communities); $\tau(v)$ represents the type of node $v$ (user, text, or community); $W_{\tau(v)} \in \mathbb{R}^{d \times d_{in}}$ is the learnable weight matrix unique to that specific node type; and $b_{\tau(v)} \in \mathbb{R}^{d}$ is the corresponding learnable bias vector. This projection ensures that features from different modalities (text vs. structural embeddings) are mapped into a shared latent space before message passing begins.

We maintain distinct, trainable weight matrices $W_{Text}$, $W_{User}$, and $W_{Community}$. This ensures that the model prevents representational collapse, maintaining the distinction between the semantic signal of a text node and the structural signal of a community node, even as they are projected into the same latent space.

\subsection{Graph Transformer Layers}

Following this projection, the projected features $H^{(0)}$ are fed into $L$ message-passing layers of an HGT. Unlike standard graph neural networks (e.g., graph convolutional networks) that share parameters across edge types, HGT maintains distinct parameters for each edge type. This allows GASTON to learn context-specific aggregation, ensuring that the pretrained community context is effectively propagated to user and text embeddings.

\subsection{Pretraining Setup and Tasks}

GASTON is pre-trained using a multi-task, self-supervised objective, similar to OMCA \cite{ma2023one}. The loss is a weighted combination of two objectives: text reconstruction and edge generation.

\begin{enumerate}
    \item \textbf{Text Reconstruction (Masked Feature Regression):} This task forces the model to learn context. We randomly mask a fraction of text nodes in each training batch, replacing their embeddings with a single, shared, learnable \texttt{mask} embedding. The HGT model must then use the surrounding graph context to predict the original, unmasked embedding. The output embedding of each masked node is passed through a decoder, and the error between the reconstructed and original embeddings is minimized. 
    
    \item \textbf{Edge Generation (Link Prediction):} This task trains the model to capture the graph structure and learn the connections among users, text, and communities. This is formulated as a contrastive link prediction objective across all edge types in our graph.

    After the HGT model produces final node embeddings for a batch, a similarity score (via dot product) is computed between relevant sets of source and target nodes. The model is then trained to distinguish positive from negative edges within the batch. We use true connections as positive samples and all other potential connections within the batch as negative samples, using cross-entropy loss to directly optimize the model to increase the similarity score for positive pairs, while decreasing it for negative pairs. This unified objective maximizes the likelihood of generating the correct graph structure, ensuring the final embeddings are structurally aware.
\end{enumerate}

The final training objective is a weighted sum of these two components: $L = \alpha \cdot L_{\text{recon}} + (1 - \alpha) \cdot L_{\text{edge}}$. Here, $\alpha$ is a hyperparameter that balances the importance of the semantic reconstruction task versus the structural link prediction task. 

\begin{figure*}[t]
  \centering
  \includegraphics[width=0.7\textwidth]{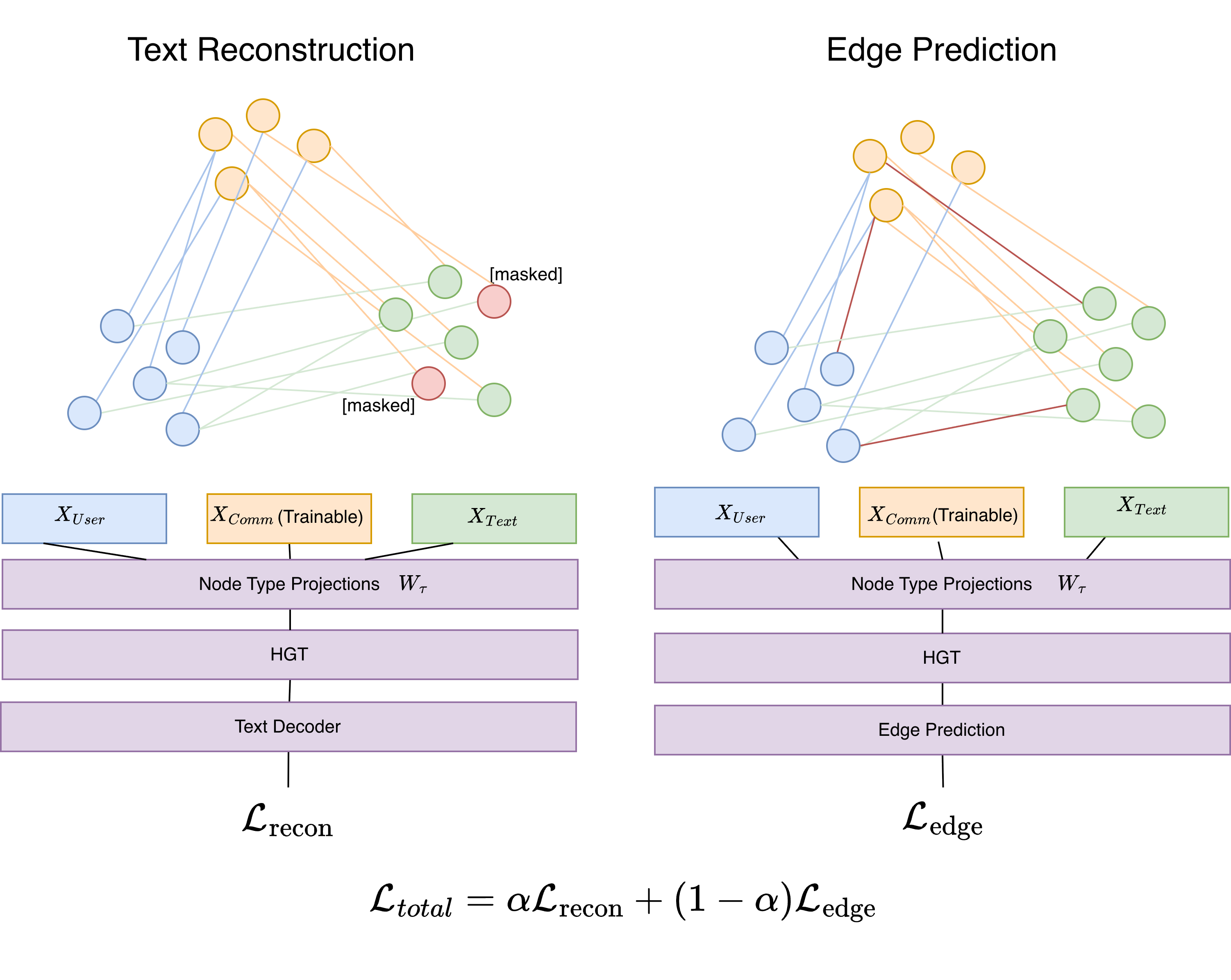}
  \caption{The GASTON pretraining architecture showing text reconstruction and edge prediction tasks.}
  \label{fig:PRE_TRAINING}
\end{figure*}

Figure \ref{fig:PRE_TRAINING} illustrates how our architectural choices drive the learning of community identity. Crucially, note that the community embedding is trainable, i.e., updated by gradients from both the text reconstruction loss ($L_{recon}$) and the edge prediction loss ($L_{edge}$). This forces the community embedding to act as a bridge: it must be structurally consistent to satisfy the edge task, but semantically informative to help reconstruct masked text in its neighbourhood.

\subsection{Pretraining Metrics}

As defined in our loss function, GASTON optimizes for both text reconstruction and link prediction. We use the following metrics to track convergence and representation quality.

For masked language modelling, we employ \textbf{Mean Squared Error (MSE)}. Given a batch of masked text nodes, we compare the reconstructed embedding $\hat{x}_i$ (output by the decoder) against the ground-truth embedding $x_i$ (from the frozen EmbeddingGemma encoder). For a set of masked nodes $M$, the loss is defined as:
\begin{equation}
    L_{recon} = \frac{1}{|M|} \sum_{i \in M} || x_i - \hat{x}_i ||^2_2
\end{equation}
Minimizing MSE ensures that HGT aggregates sufficient neighbourhood context to accurately recover a post's semantic content without seeing its text.

For link prediction, we monitor performance on all forward edge types: $\langle Text, posted\_by, User \rangle$, $\langle Text, posted\_in, Community \rangle$, and $\langle User, active\_in, Community \rangle$. We evaluate the model using \textbf{Cross Entropy Loss}.  As our edge generation objective treats link prediction as a binary classification problem (distinguishing true edges from negative samples), this measures how well the model discriminates between valid and invalid structural connections.

We compute the loss separately for each edge type to ensure that the model learns all relationships effectively. For a given batch of $N$ edges, where $y_i \in \{0,1\}$ represents the true label (1 for real edges, 0 for negative samples) and $\hat{y}_i$ represents the model's predicted probability, the loss is calculated as:

\begin{equation}
    L_{edge} = - \frac{1}{N} \sum_{i=1}^{N} [y_i \log(\hat{y}_i) + (1 - y_i) \log(1 - \hat{y}_i)]
\end{equation}

Minimizing this cross-entropy loss indicates that the learned user and community embeddings are aligned in the latent space, maximizing the likelihood of observing the true graph structure while suppressing random pairings.

\section{Experiments}

\subsection{Pretraining Data}

For pretraining GASTON, we use the Reddit historical archive, acquired via the Reddit API, consisting of all available user, community, and post data within the selected time range. To ensure heterogeneity in topics and community dynamics, a stratified sampling strategy was employed. We curated a representative subset of subreddits using data collected between January 2017 and December 2021. This subset encompasses a broad range of topics, from highly technical discussions to general-interest conversations.

To improve data quality and distill meaningful signals from the content, two primary noise reduction heuristics were applied. First, only top-level submissions that received a score of at least 25 were selected. This score-based filter serves as a community-driven signal of quality, as posts with higher scores are more likely to contain substantive content and less likely to be spam or low-effort submissions.

Second, conversational threads were truncated to a depth of four comments. This pruning strategy removes the ``long-tail'' of discussion, which often consists of off-topic remarks, personal exchanges, or repetitive content. By focusing on the initial, high-quality portions of the threads, this step reduces conversational noise while retaining the core exchange.

These curation and noise-reduction techniques yield a pretraining dataset (heterogeneous graph) with 6.4 million pieces of text, 6132 communities, 4.7 million users, 6.4
million $\langle Text, posted by, User\rangle$ edges, 6.4 million $\langle Text, posted in, Community\rangle$ edges, and 21 million $\langle User, active in, Community\rangle$ edges.

\subsection{Fine-Tuning for Downstream Tasks}

\begin{table*}
\centering
\caption{Summary of downstream evaluation tasks}
\begin{tabular}{l l l l l l}
\toprule
\textbf{Task} & \textbf{Dataset} & \textbf{Size} & \textbf{Type} & \textbf{Level} & \textbf{Metrics} \\
\midrule
Mental Health Detection & Dreaddit & 4k & Binary Classif. & Node (Text) & Acc, F1 \\
Toxicity Scoring & Ruddit & 5k & Regression & Node (Text) & RMSE, $r$ \\
Norm Violation & NormVio & 45k & Binary Classif. & Node (Text) & Acc, F1 \\
Hate Speech Detection & HatefulDisc. & 61k & Binary Classif. & Node (Text) & Acc, F1 \\
Comm. Recommendation & Reddit (Ours) & 9.8M & Link Prediction & Edge & NDCG, MRR \\
\bottomrule
\end{tabular}
\label{tab:downstream_tasks}
\end{table*}

Table \ref{tab:downstream_tasks} summarizes the downstream tasks, selected to demonstrate that the community and user context learned by GASTON is critical when disambiguating the meaning of online text, especially in socially nuanced settings.  The first four are classification or regression tasks on individual Reddit posts and comments, for which we fine-tune simple classification or regression heads atop GASTON, using the text node embedding as input. The corresponding datasets only contain posts and their labels or scores. Therefore, to produce complete fine-tuning datasets, we use the Reddit API to retrieve the corresponding user and community data, transforming the posts into heterogeneous graphs required by GASTON. These four tasks are as follows.
\begin{enumerate}
\item Mental health detection using the Dreaddit dataset: a binary classification problem to predict whether a post from a mental health subreddit expresses stress or no stress \cite{turcan-mckeown-2019-dreaddit}.
\item Toxicity scoring using the Ruddit dataset: a regression problem to predict the offensiveness score of posts, ranging from {-1} (maximally supportive) to {+1} (maximally offensive) \cite{hendrickx2021ruddit}.
\item Norm violation detection using the NormVio dataset: a binary classification problem to detect whether a comment violates its subreddit's rules  \cite{park2021detecting}. This dataset moves beyond toxicity to encompass a variety of rules enforced by moderators, including format, off-topic content, and incivility. 
\item Hate speech detection using the HatefulDiscussions dataset: a binary classification problem to predict whether a comment is hateful or not \cite{hebert2023qualitative}.
\end{enumerate}

Finally, we consider community recommendation, formulated as an edge-prediction problem. This task requires the model to predict novel yet likely edges between user nodes and community nodes, thereby predicting which communities a specific user is likely to engage with. The fine-tuning process is structured as a standard edge-prediction task: we train a prediction head to estimate the probability of an edge between the user node embedding and the community node embedding. During evaluation, the goal is to distinguish between a set of true (observed) and false (unobserved) edges. We fine-tune the model on user-community interactions completely disjoint from the pretraining set. 

\subsection{Evaluation Metrics}
\label{sec:evaluation_metrics}

For node-level classification tasks (Dreaddit, NormVio, HatefulDiscussions), we report \textbf{Accuracy} and \textbf{Macro-F1}, the latter calculated as the harmonic mean of Precision ($P$), the fraction of relevant instances among the retrieved instances, and Recall ($R$), the fraction of relevant instances that were retrieved. While accuracy measures overall correctness, Macro-F1 is critical for our datasets, which often exhibit class imbalance.

For the Ruddit toxicity scoring task, which requires predicting a continuous value, we evaluate performance using \textbf{Root Mean Squared Error (RMSE)} and \textbf{Pearson Correlation Coefficient ($r$)}. To calculate RMSE, we first compute the Mean Squared Error (MSE), the average squared difference between the predicted toxicity score $\hat{y}$ and the ground truth score $y$. For a dataset of size $N$, MSE is defined as:
\begin{equation}
    MSE = \frac{1}{N} \sum_{i=1}^{N} (y_i - \hat{y}_i)^2
\end{equation}
We report the RMSE, which is the square root of the MSE:
\begin{equation}
    RMSE = \sqrt{MSE}
\end{equation}
Using RMSE returns the error metric to the original scale of the dataset (toxicity scores ranging from -1 to 1), allowing for a direct interpretability of how far, on average, the model's predictions deviate from human annotations. Additionally, Pearson's $r$ measures the linear correlation between predicted and ground-truth toxicity, indicating how well the model ranks toxicity relative to human annotators.

For the user-community recommendation task, which is a ranking problem, we adopt information retrieval metrics: \textbf{Mean Reciprocal Rank (MRR)} and \textbf{Normalized Discounted Cumulative Gain (NDCG)}.
MRR evaluates the model's ability to rank the correct community high in the predicted list. For a set of users $U$, where $rank_u$ is the position of the ground-truth community in the model's sorted prediction list:
\begin{equation}
    MRR = \frac{1}{|U|} \sum_{u \in U} \frac{1}{rank_{u}}
\end{equation}
NDCG@$k$ (specifically at $k=10$) evaluates the quality of the top-$k$ recommendations, placing greater weight on correct predictions at the top of the list. NDCG@$k$ is calculated as the Discounted Cumulative Gain (DCG) divided by the Ideal DCG (IDCG). DCG measures the quality of the ranking by summing the relevance scores of recommendations, discounted logarithmically by their position in the list. IDCG represents the score of a perfect ranking order.

\subsection{Baselines}

We evaluate three baselines (recall Figure~\ref{fig:model_comparison}). First, BERT-only serves as a text-only benchmark; it is a standard pretrained BERT classifier fine-tuned on the post's text. Second, BERT + Community represents a simple method of incorporating social context. In this model, we employ community2vec  \cite{martin2017community2vec} to generate user-membership-based community embeddings. During fine-tuning, this community vector is concatenated with the BERT text representation before being passed to the classification head.  Finally, OMCA is a state-of-the-art model that unifies users, text, and communities.  This baseline allows us to test whether GASTON's structural contrastive initialization and modern text-embedding backbone improve upon the current standard in unified community modelling.  

To ensure a rigorous comparison, we independently replicated the OMCA architecture. As the official implementation was unavailable, we rebuilt the model and strictly adhered to the described methodology, using XLNet \cite{yang2019xlnet} as the text encoder.  In our replication of OMCA, the final text node embedding is derived by applying mean pooling over the last hidden state of the XLNet outputs. Formally, given a sequence of token vectors $H = \{h_1, ..., h_n\}$, the text representation $x_{text}$ is calculated as the element-wise average: $x_{text} = \frac{1}{n} \sum_{i=1}^{n} h_i$. This text-centric approach is then extended to initialize the social nodes in the OMCA baseline: the initial embedding for a user node is computed as the mean of all text embeddings they have authored, and a community node's initial embedding is the mean of all text embeddings posted within it.

\subsection{Experimental Environment}

Experiments were conducted using an RTX 6000 Ada Generation GPU with 160 GB of RAM and 16 CPU cores. All models were trained and evaluated using PyTorch, PyTorch Geometric, and PyTorch-Lightning. All models share a consistent hyperparameter configuration. The embedding dimension for all nodes and the hidden channel size within the HGT layers were set to $d=768$. The HGT architecture consists of three layers, each employing 8 attention heads. In total, the full GASTON model contains approximately 26 million trainable parameters.  

For pretraining, we utilized the Adam optimizer with a learning rate of $1\mathrm{e}{-4}$ and a batch size of 128. To handle the scale of the full heterogeneous graph, which exceeds typical GPU memory constraints, we employed a \texttt{NeighborLoader} to sample local subgraphs for each batch. The loss-weighting parameter $\alpha$, which balances the text reconstruction and edge-generation objectives, was set to $0.5$ to assign equal importance to both structural and semantic learning signals.

Each pretrained model was trained for six epochs to generate the parameters used for fine-tuning. This procedure is computationally intensive due to the high memory footprint of the graph states; pre-training the fully trainable GASTON model required approximately 34 hours on the specified hardware.

The majority of our downstream tasks have a predefined train-test split. For the remaining datasets, we use an 80-10-10 train-test-validation split. We perform train, test, and validation dataset splitting at the user level to prevent data leakage during training. For downstream tasks, we use the provided train/test split where applicable and set aside 15\% of the training data for validation. To address dataset imbalance in downstream-task data, we assign weights to each class when computing the cross-entropy loss. We apply seeds consistently across all models and data splits to ensure reproducibility and a fair comparison across experiments.

\subsection{GASTON Performance Summary}

\begin{table*}
\centering
\caption{Accuracy and Macro-F1 for classification, Pearson's r and RMSE for regression, and NDCG@10 and MRR@10 for recommendation. \textbf{Higher is better} for all metrics, except \textbf{RMSE} where lower is better ($\downarrow$). Best-performing model is in bold.}
\small
\begin{tabular}{l|cc|cc|cc|cc|cc}
\toprule
\multirow{3}{*}{\textbf{Model}} & \multicolumn{8}{c|}{\textbf{Node-Level Tasks}} & \multicolumn{2}{c}{\textbf{Edge-Level Task}} \\
\cmidrule{2-11}
 & \multicolumn{2}{c|}{\textbf{Dreaddit}} & \multicolumn{2}{c|}{\textbf{NormVio}} & \multicolumn{2}{c|}{\textbf{HatefulDisc.}} & \multicolumn{2}{c|}{\textbf{Ruddit}} & \multicolumn{2}{c}{\textbf{Recommend.}} \\
 & F1 & Acc & F1 & Acc & F1 & Acc & $r$ & RMSE ($\downarrow$) & NDCG@10 & MRR@10 \\
\midrule
BERT-only & 0.911 & 0.854 & 0.696 & 0.754 & 0.841 & 0.828 & 0.663 & 0.250 & - & - \\
BERT + Community & \textbf{0.939} & \textbf{0.894} & 0.710 & 0.761 & 0.889 & 0.854 & \textbf{0.670} & \textbf{0.248} & - & - \\
\midrule
OMCA (Ma et al.) & 0.929 & 0.878 & 0.961 & 0.953 & 0.916 & 0.892 & 0.525 & 0.292 & 0.011 & 0.009 \\
\midrule
\textbf{GASTON (Ours)} & 0.914 & 0.856 & \textbf{0.971} & \textbf{0.967} & \textbf{0.920} & \textbf{0.901} & 0.598 & 0.271 & \textbf{0.055} & \textbf{0.061} \\
\bottomrule
\end{tabular}
\label{tab:main-results}
\end{table*}

Table \ref{tab:main-results} presents the main experimental results. GASTON achieves state-of-the-art performance on three out of the five downstream tasks, excelling in socially nuanced classification problems. Notably, in the Norm Violation and Hate Speech Detection tasks, GASTON significantly outperforms both the text-only baselines and the OMCA graph baseline (+21\% F1 improvement over BERT on NormVio). This confirms that community-aware representations are critical for detecting context-dependent violations. In the community recommendation, GASTON also achieves the highest ranking performance, validating our contrastive initialization.

However, for tasks with smaller datasets such as Ruddit (toxicity) and Dreaddit (stress), the simpler BERT baselines perform competitively with, or slightly better than, the graph models. As detailed below, we attribute this to the risk of over-parameterization when applying graph neural networks to data-scarce environments. However, even in these cases, the BERT + Community baseline outperforms the text-only BERT baseline (e.g., 0.894 vs 0.854 Accuracy on Dreaddit). Thus, regardless of model complexity, incorporating community context is essential for accurate social prediction.

\subsection{GASTON Performance Analysis}

Next, we analyze specific failure and success cases. We find that GASTON captures context-dependent nuances, such as distinguishing support from toxicity, that text-only baselines systematically miss. This suggests that the model's underlying representations are better aligned with the complex socio-semantics of online interaction.

\subsubsection*{The Challenge of Small Datasets: Ruddit}
While GASTON outperforms OMCA on the Ruddit toxicity regression task ($0.271$ vs $0.292$ RMSE), the simpler BERT-based baselines achieve slightly lower RMSE ($0.248$). This highlights the over-parameterization risk inherent in graph neural networks when applied to small datasets. Ruddit contains fewer than 6,000 samples. In a data-scarce task, the structural signal in the graph is sparse, and the model struggles to propagate meaningful information without overfitting. The BERT-only baseline, which relies solely on the text without the overhead of learning graph topology, is less prone to this noise on smaller datasets. This suggests that while GASTON excels at scale, its graph-aware components require a critical mass of data to distinguish valid social signals from noise.

\subsubsection*{The Challenge of Small Datasets: Dreaddit}
We observe a similar trend with the Dreaddit dataset. GASTON achieved an accuracy of $0.856$, which is effectively the same as the basic BERT baseline ($0.854$) but slightly lower than the simpler BERT + Community method ($0.894$). With only about 4,000 nodes in the training graph, the dataset may be too small for a graph neural network to be effective. While GASTON attempts to learn deep structural relationships, there may be insufficient interaction data to generate a strong signal. In this case, the simpler baseline, which adds a static community embedding to the text, provides useful context without requiring complex training.

\subsubsection*{Sparsity in Recommendation Tasks}
On the community recommendation task, GASTON significantly outperforms OMCA (NDCG@10 of $0.055$ vs $0.010$). However, the absolute values for these metrics remain low due to the extreme sparsity of the user-community interaction graph. The average user in the validation set has interacted with only 4.7 communities, with a median of two.  Furthermore, the evaluation metrics (NDCG and MRR) penalize the model for ``incorrect'' ranking order. However, the ground truth is binary (interaction vs. non-interaction) rather than an ordered preference list. The model might recommend a relevant community that the user has not yet visited; in the evaluation, this counts as a false positive. If the dataset contained richer user histories, we hypothesize that the gap between GASTON and the baselines would widen further, as the collaborative filtering signal would be stronger.

\subsubsection*{Case Study: Contextual Norm Violation} The most significant performance gain is observed in the Norm Violation (NormVio) task, where GASTON achieves an F1 score of $0.972$ compared to the BERT baseline's $0.696$. This improvement confirms that what constitutes a violation is defined by the community, not just the text.

\subsubsection*{Case Study: The Necessity of Community Context}

We highlight two examples from our test set in which graph-based models (GASTON and OMCA) succeeded, whereas the text-only baseline failed.

\textbf{Example 1: Sharing Personal Information}
\begin{quote}
    \textit{``Can you add my snap? D***********103 I'm 18''}
\end{quote}
\textbf{Prediction Analysis:}
\begin{itemize}
    \item \textbf{BERT-only} classifies this as \textit{Non-Violation}. The model analyzes the semantics and finds no toxicity, aggression, or hate speech. It interprets the text as a neutral social request.
    \item \textbf{OMCA} classifies this as \textit{Non-Violation}. Despite using a graph structure, the model fails to capture the specific constraints of \texttt{r/teenagers}. It prioritizes the benign semantic content of the post over the signal indicating a rule violation.
    \item \textbf{GASTON} correctly classifies this as a \textit{Violation}. The graph models incorporate the \texttt{r/teenagers} community node, which represents a subreddit with strict safety rules against sharing Personally Identifiable Information (PII) or soliciting off-platform contact. By leveraging the community context, the models understand that exchanging Snapchat handles is a violation in this specific space.
\end{itemize}

\textbf{Example 2: Unconstructive Commentary}
\begin{quote}
    \textit{``You're pretty enough to get a boyfriend? You're ahead of the rest of us.''}
\end{quote}
\textbf{Prediction Analysis:}
\begin{itemize}
    \item \textbf{BERT-only} classifies this as \textit{Non-Violation}. The text contains positive words such as ``pretty'' and ``ahead'', which can confuse a model trained primarily on toxicity or sentiment. It likely interprets this as a compliment.
    \item \textbf{Graph Models (GASTON \& OMCA)} correctly classify this as a \textit{Violation}. In the context of this being in the BodyDysmorphia subreddit, this comment functions as a backhanded compliment or unconstructive comparison, violating the norm ``Be Kind''. The graph models perceive the misalignment between this type of post and the community's norms for supportive interaction.
\end{itemize}

These examples demonstrate that text-only models suffer from context blindness, and existing graph models fail to capture the nuance of these pieces of text. They fail to detect violations that are defined by rules rather than by vocabulary. GASTON provides the grounding to align the definition of a violation with the specific social environment.

\subsubsection*{Case Study: The Echo Chamber Effect in Extreme Communities}
A failure mode of models such as OMCA is the ``normalization of toxicity'' within echo chambers. If a community's discourse is uniformly extreme, the community's embedding will reflect this extremism. A hateful comment might be mathematically close to the community's center, leading the model to interpret it as acceptable.

We observe this phenomenon in \texttt{r/TumblrInAction}, a subreddit known for extreme anti-progressive rhetoric, which was subsequently banned from Reddit for violating its hate speech policies.  Consider the following examples from this community, which GASTON correctly identified as Hateful, but OMCA misclassified as Non-Hateful:
\begin{itemize}
    \item \textit{``The act of blowing a tr*nny takes on very, very different meanings depending on the context...''}
    \item \textit{``ehh... it gets iffy. Some people aren't cool with it even if they have a vagina. [...] If it happened to me I'd be hurt maybe a little offended, but overall I would understand and not blame the [sic]. (unless they went full-on tr*nny hate)''}
    \item \textit{``She's pretty well-known for being volatile [...]. Screaming at and berating the ref for doing his job [...] shows an astonishing lack of class. She was acting like a selfish brat and ruined the actual winner's big moment by being such a tw*t. I'm appalled anyone is defending her behavior.''}
\end{itemize}

\textbf{Failure Analysis:}
\begin{itemize}
    \item \textbf{OMCA} initializes the community embedding by averaging the posts within it. In a subreddit such as \texttt{r/TumblrInAction}, where hostility was the norm, these comments align with the average discourse. The model likely perceived these comments as conforming to community standards and classified them as non-hateful.
    \item \textbf{GASTON} learns community identity through user co-occurrence. The users active in \texttt{r/TumblrInAction} likely overlapped with those in other controversial or toxic communities. GASTON learned a community embedding that placed \texttt{r/TumblrInAction} in a high-risk region of the latent space. Consequently, when the model processed these comments, the \textit{Community Node} served as a bias-correction mechanism, or contextual prior,
    informing the classifier that the text originated from a known source of hate speech.
    Crucially, this does not mean that GASTON labels all speech in this community as hateful. Rather, for ambiguous or borderline content, which might pass as neutral in isolation, the model uses this negative community signal to shift the decision boundary, correctly identifying the hateful intent that text-only models miss.
\end{itemize}

This provides evidence that GASTON is robust to the ``Echo Chamber'' effect, capable of detecting toxicity even when it is the dominant norm within a specific subgroup.

\subsection{Ablations}

To isolate the contributions of specific components within GASTON, we evaluate three variations of our architecture:

\begin{enumerate}
    \item \textbf{GASTON (Frozen Embeddings)}: We initialize community nodes using our contrastive (BPR) method, but freeze these weights during the graph pretraining phase. This tests whether the structural signal learned from user overlap is sufficient, or whether community representations benefit from updates based on semantic signals during pretraining.
    \item \textbf{GASTON (Average Initialization):} We remove our contrastive initialization and replace it with the text-averaging strategy used by OMCA (initializing a community as the mean of its posts). By keeping the rest of the GASTON architecture unchanged, this ablation isolates the specific impact of modelling communities based on structural versus semantic features.
    \item \textbf{GASTON (with XLNet):} We replace our text encoder, EmbeddingGemma, with XLNet, matching the encoder used in OMCA. This allows us to disentangle the performance gains attributed to our graph architecture from those attributed simply to using a more modern, instruction-tuned language model.
\end{enumerate}

\begin{table*} 
\centering
\caption{The impact of the language model (Gemma vs. XLNet) and the community initialization strategy (Trainable Community vs. Frozen vs. Average Initialization). \textbf{Higher is better} for all metrics, except \textbf{RMSE} where lower is better ($\downarrow$). Recommendation metrics (NDCG and MRR) are reported at k=10.}
\small
\begin{tabular}{l|cc|cc|cc|cc|cc}
\toprule
\multirow{3}{*}{\textbf{Model Variant}} & \multicolumn{8}{c|}{\textbf{Node-Level Tasks}} & \multicolumn{2}{c}{\textbf{Edge-Level Task}} \\
\cmidrule{2-11}
 & \multicolumn{2}{c|}{\textbf{Dreaddit}} & \multicolumn{2}{c|}{\textbf{NormVio}} & \multicolumn{2}{c|}{\textbf{HatefulDisc.}} & \multicolumn{2}{c|}{\textbf{Ruddit}} & \multicolumn{2}{c}{\textbf{Recommend.}} \\
 & F1 & Acc & F1 & Acc & F1 & Acc & $r$ & RMSE ($\downarrow$) & NDCG & MRR \\
\midrule
\textbf{GASTON (Ours)} & \textbf{0.914} & \textbf{0.856} & 0.971 & 0.967 & \textbf{0.920} & \textbf{0.901} & \textbf{0.598} & \textbf{0.271} & 0.055 & 0.061 \\
\midrule
\multicolumn{11}{l}{\textit{Community Initialization Ablations}} \\
\quad -- with Frozen Embeddings & 0.888 & 0.816 & \textbf{0.972} & 0.977 & 0.896 & 0.873 & 0.422 & 0.304 & \textbf{0.057} & 0.066 \\
\quad -- with Average Init & 0.906 & 0.842 & 0.754 & 0.783 & 0.885 & 0.861 & 0.386 & 0.312 & 0.052 & \textbf{0.069} \\
\midrule
\multicolumn{11}{l}{\textit{Language Model Ablation}} \\
\quad -- with XLNet (instead of Gemma) & 0.889 & 0.817 & 0.971 & \textbf{0.978} & 0.913 & 0.895 & 0.561 & 0.277 & 0.033 & 0.043 \\
\bottomrule
\end{tabular}
\label{tab:ablation-study}
\end{table*}

\subsubsection{Impact of Language Model: XLNet vs. EmbeddingGemma}

As shown in Table \ref{tab:ablation-study}, replacing XLNet \cite{yang2019xlnet} with EmbeddingGemma \cite{google2024embeddinggemma} yields performance gains across all tasks, most notably in Ruddit and recommendation. We attribute this improvement to the difference in training objectives between the two models. XLNet is a generalized autoregressive pretraining method designed primarily for sequence generation and probability estimation. While effective for many NLP tasks, it is not explicitly optimized to produce semantically meaningful embeddings of full sentences without fine-tuning.  In contrast, EmbeddingGemma is an encoder-only model that has been instruction-tuned for retrieval and representation tasks. 

\subsubsection{Community Initialization Strategy: Trainable vs. Frozen Weights}

A core contribution of GASTON is the use of collaboratively learned community nodes that are initialized using a contrastive BPR objective, then fine-tuned during pretraining. In this ablation study, we investigate whether these initialized weights should remain fixed (serving as a static structural prior) or be trainable (allowing them to evolve).

\textbf{Frozen Embeddings:} When we freeze the community embeddings after the contrastive initialization step, the model still outperforms the Average Initialization baseline significantly (e.g., NormVio F1 of $0.971$ vs $0.754$). This validates the structural ``signature'' learned from user membership patterns: knowing \textit{who} interacts in a community is more informative than simply averaging \textit{what} they write.

\textbf{Trainable Embeddings:} However, the fully trainable approach achieves the best overall performance. By allowing the initialized weights to be updated during graph pretraining, GASTON benefits from both structural and semantic information. The BPR initialization provides a strong prior belief about which communities are related based on user overlap. The graph training then refines this position based on the text content flowing through the network. This allows the community embedding to evolve into a hybrid representation: it starts as a purely structural entity (defined by its user base) and becomes a socio-semantic entity (defined by its users \textit{and} its evolving norms), providing the context necessary for nuanced tasks like norm violation detection.

\subsection{Embedding Visualization}

\begin{figure}
  \centering
  \includegraphics[width=\columnwidth]{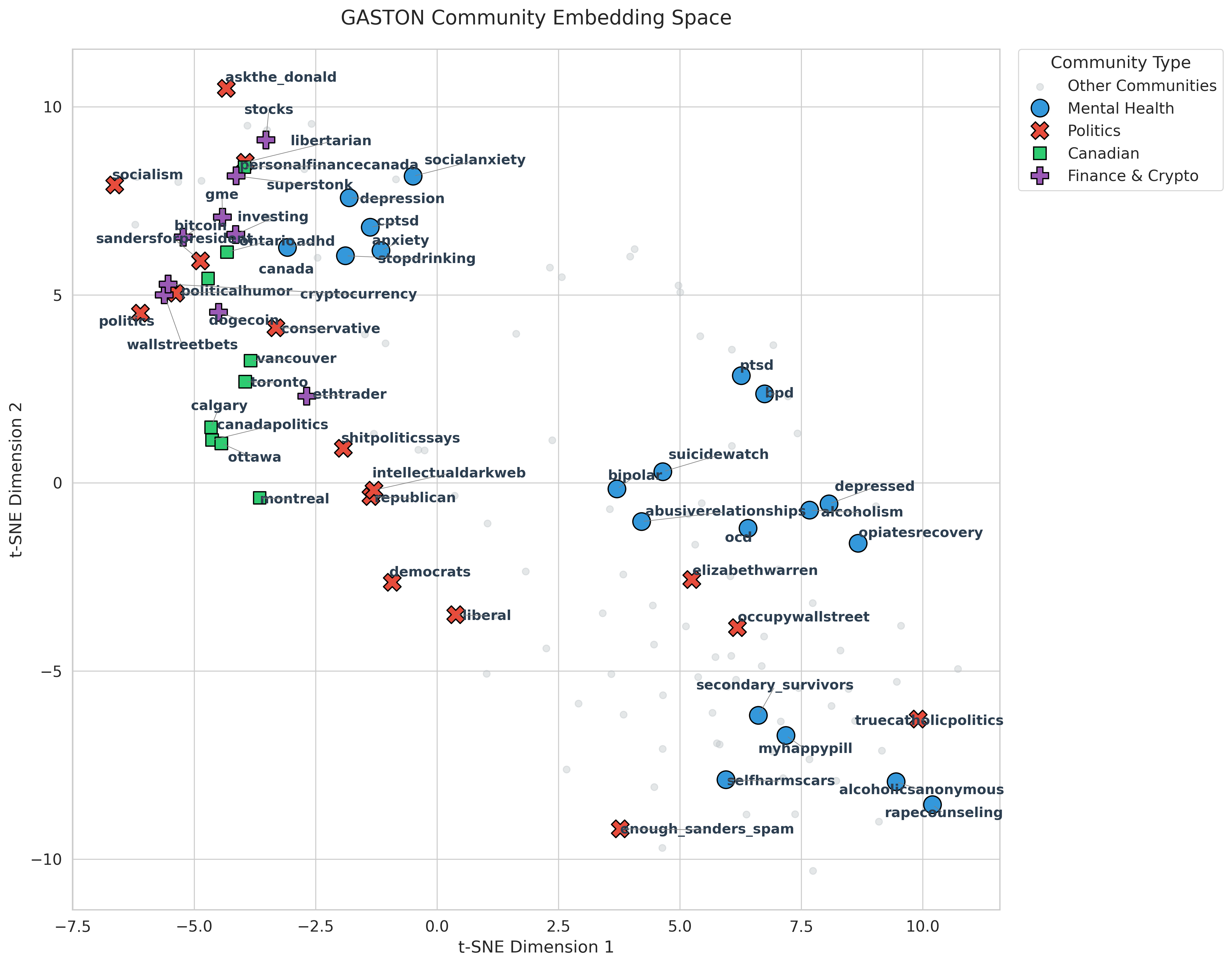}
  \caption{Visualizing GASTON's community embeddings}
  \label{fig:comm_embeddings}
\end{figure}

Finally, to validate the structure of our learned representations, we project a sample of the final community embeddings into a 2D space using t-SNE (t-Distributed Stochastic Neighbour Embedding), a method for visualizing high-dimensional data by assigning each data point a location in a 2D map. Our community-embedding visualization in Figure \ref{fig:comm_embeddings} shows that GASTON organizes the latent space into semantically coherent clusters. We observe tight groupings of communities based on shared topics and user bases:

\begin{itemize}
    \item \textbf{Mental Health:} Communities such as \texttt{r/depression}, \texttt{r/suicidewatch}, and \texttt{r/anxiety} form a cluster, reflecting the high user overlap and shared support function of these spaces.
    \item \textbf{Politics:} Political subreddits, including \texttt{r/democrats}, \texttt{r/republican}, and \texttt{r/politics}, cluster together. Despite their ideological differences, the model places them in the same structural neighbourhood, capturing their shared utility as forums for political discourse.
    \item \textbf{Canadian:} Subreddits tailored specifically to Canadians clearly form a cluster in the top left quadrant of the embedding space, showing their contextual similarity amongst other subreddits.
    \item \textbf{Finance \& Crypto:} A distinct cluster emerges for financial interests, grouping \texttt{r/investing} with \texttt{r/bitcoin} and \texttt{r/wallstreetbets}.
\end{itemize}

This geometric organization demonstrates that GASTON has translated user interaction patterns and semantic signals into meaningful proximity, where the distance between two community vectors serves as a proxy for their sociological similarity. Crucially, this structure emerges without manual curation. Unlike previous work \cite{waller2021quantifying} that mapped the social landscape of Reddit by explicitly projecting communities onto axes defined by researcher-selected poles, GASTON enables latent sociological dimensions to surface entirely from the user-community and semantic topology of the graph.

\section{Conclusions and Implications}

We presented GASTON, a heterogeneous graph architecture that integrates the semantic power of language models with the structural insights of graph learning. We introduced a contrastive initialization technique that treats communities as entities and fine-tunes them during pretraining to generate meaningful contextual representations for downstream tasks. Our evaluation confirms that this approach yields promising representations. 

Our quantitative results demonstrate GASTON's performance across standard metrics. However, standard metrics do not fully capture a model's societal impact. With respect to social implications, we argue that, rather than developing enhanced censorship capabilities, we should use community-aware representations to empower users with the transparency they need to navigate online spaces safely. 
Our efforts complement works such as \citet{zhang2017community} which also seeks to further methods in understanding engagement patterns. 
GASTON can enable this vision of community-aware analysis in two ways.

First, research indicates that while online communities provide vital peer support and identity affirmation for marginalized groups, they also carry significant risks. These include exposure to toxicity, misinformation, and ``problematic use'' patterns that correlate with anxiety and depression \cite{rideout2018digital, primack2017social}. A critical failure of standard text-only models in this domain is their tendency to flatten this complexity. By treating all negative text as toxic, context-agnostic models risk dismantling positive support networks. For example, a user venting about distress in a recovery group might be flagged for removal, stripping them of their support system when they are most vulnerable. In contrast, GASTON's contextual embeddings would be more likely to capture these local differences.

Furthermore, while support is vital, online communities can also facilitate the formation of echo chambers that normalize distress and amplify emotional contagion, the phenomenon in which users unconsciously converge on the emotional states of their peers \cite{kramer2014experimental}. For vulnerable individuals, the risk includes exposure to explicit abuse (which standard models can catch), but also exposure to \textit{maladaptive norms}. A community that subtly encourages rumination or normalizes self-destructive behaviour can trap users in a negative feedback loop, often without their realizing it. Traditional context-agnostic models fail to detect this risk because they analyze posts in isolation, missing the normative pressure of the environment. GASTON addresses this gap. By learning a distinct representation for the community itself, our model captures the emotional baseline of a digital space, distinguishing between a support group that fosters recovery and an echo chamber that reinforces pathology. This enables a shift from moderation to user agency, providing the building blocks of an ``algorithmic nutrition label'' \cite{kelley2009standardizing} that enables users to assess the health of a community before engaging.

\section{Limitations and Future Work}

In future work, we plan to enrich the graph representation of social interactions by adding more granular edge types, such as $\langle text, replying\_to, text \rangle$ relations. Modelling the reply-tree structure of conversations reveals even more conversational context, distinguishing between top-level comments and deep, branching discussions. Furthermore, the nodes could be enriched with additional features, such as user activity frequency or community size. 

Additionally, social media is an inherently multimodal environment, yet GASTON currently processes text while ignoring images and videos in posts. A useful extension would be to incorporate the visual modality associated with posts. This would involve transforming GASTON into a Multimodal Graph Learning (MGL) framework, similar to architectures proposed in recommendation domains \cite{Wei2019MMGCN}. In such a model, a text node's initial representation would be a fused embedding derived from its text and image content. This may enable the model to develop a deeper understanding of content that relies on the interplay between text and image, such as memes, infographics, or visual based hate speech.

\section{Ethical Statement}

All data used in this research were previously publicly released, and when using and accessing Reddit data, we followed the Reddit Terms of Service\footnote{The Reddit Developer Terms of Service can be found at \url{https://redditinc.com/policies/developer-terms}}. We also ensured that all data collected from Reddit was anonymized prior to use.

\section{Acknowledgements}
Hebert gratefully thanks the financial support from the Natural Sciences and Engineering Research Council of Canada (NSERC) through the Vanier Graduate Scholarship, as well as from the IEEE Nick Cercone Graduate Scholarship. The authors also thank the University of Waterloo for its institutional support.

\bibliography{aaai2026}


\pagebreak

\section{Paper Checklist}

\begin{enumerate}


\item For most authors...
\begin{enumerate}
  \item  Would answering this research question advance science without violating social contracts, such as violating privacy norms, perpetuating unfair profiling, exacerbating the socio-economic divide, or implying disrespect to societies or cultures?
    \answerYes{Yes. Answering this research question advances our understanding of online social dynamics using publicly available, anonymized data, and we ensure data quality and representation without violating social contracts.}
  \item Do your main claims in the abstract and introduction accurately reflect the paper's contributions and scope?
    \answerYes{Yes. Claims made regarding GASTON's ability to model community context made in the Abstract and Introduction are addressed in the Experiments, GASTON Performance Summary and GASTON Performance Analysis sections.}
  \item Do you clarify how the proposed methodological approach is appropriate for the claims made? 
    \answerYes{Yes. We justify the use of a heterogeneous graph transformer and contrastive initialization in the Our Solution section to address the specific limitations of text-only and structure-only models identified in our problem statement.}
  \item Do you clarify what are possible artifacts in the data used, given population-specific distributions?
    \answerYes{Yes. We discuss the data collection, stratification, and noise reduction strategies used in the Pretraining Data subsection of our Experiments section to mitigate any potential artifacts.}
  \item Did you describe the limitations of your work?
    \answerYes{Yes. We describe limitations, such as the model's performance on small datasets in the GASTON Performance Analysis section, and limitations that suggest future work in the Limitations and Future Work section.}
  \item Did you discuss any potential negative societal impacts of your work?
    \answerYes{Yes. We discuss the potential negative societal impacts such as the risk of moderation in the Conclusions and Implications section.}
  \item Did you discuss any potential misuse of your work?
    \answerYes{Yes. We address the potential for misuse in the Conclusions and Implications section, specifically arguing against using these capabilities for enhanced censorship and instead advocating for user empowerment.}
  \item Did you describe steps taken to prevent or mitigate potential negative outcomes of the research, such as data and model documentation, data anonymization, responsible release, access control, and the reproducibility of findings?
    \answerYes{Yes. We mitigate potential negative outcomes by providing open-sourced code for transparency, ensuring all hyperparameters are available in the Experimental Environment section, and strictly using anonymized, public data as stated in the Ethical Statement.}
  \item Have you read the ethics review guidelines and ensured that your paper conforms to them?
    \answerYes{Yes, we have reviewed the ethics guidelines and ensured our paper conforms to them.}
\end{enumerate}

\item Additionally, if your study involves hypotheses testing...
\begin{enumerate}
  \item Did you clearly state the assumptions underlying all theoretical results?
    \answerNA{NA}
  \item Have you provided justifications for all theoretical results?
    \answerNA{NA}
  \item Did you discuss competing hypotheses or theories that might challenge or complement your theoretical results?
    \answerNA{NA}
  \item Have you considered alternative mechanisms or explanations that might account for the same outcomes observed in your study?
    \answerNA{NA}
  \item Did you address potential biases or limitations in your theoretical framework?
    \answerNA{NA}
  \item Have you related your theoretical results to the existing literature in social science?
    \answerNA{NA}
  \item Did you discuss the implications of your theoretical results for policy, practice, or further research in the social science domain?
    \answerNA{NA}
\end{enumerate}

\item Additionally, if you are including theoretical proofs...
\begin{enumerate}
  \item Did you state the full set of assumptions of all theoretical results?
    \answerNA{NA}
  \item Did you include complete proofs of all theoretical results?
    \answerNA{NA}
\end{enumerate}

\item Additionally, if you ran machine learning experiments...
\begin{enumerate}
  \item Did you include the code, data, and instructions needed to reproduce the main experimental results (either in the supplemental material or as a URL)?
    \answerYes{Yes. We include an anonymous URL to our code repository in the Code section on the first page, and we use publicly available datasets for pretraining and downstream tasks as detailed in the Experiments section.}
  \item Did you specify all the training details (e.g., data splits, hyperparameters, how they were chosen)?
    \answerYes{Yes. All training details, including learning rates, batch sizes, data splits, and our hyperparameter configuration, are specified in the Experimental Environment section.}
  \item Did you report error bars (e.g., with respect to the random seed after running experiments multiple times)?
    \answerNo{No. Due to the high computational costs associated with pretraining large-scale heterogeneous graph transformers, we do not report error bars, which is consistent with the methodology of prior work in this domain.}
  \item Did you include the total amount of compute and the type of resources used (e.g., type of GPUs, internal cluster, or cloud provider)?
    \answerYes{Yes. We detail the computational resources in the Experimental Environment section.}
  \item Do you justify how the proposed evaluation is sufficient and appropriate to the claims made? 
    \answerYes{Yes. We justify our evaluation by testing on five downstream tasks (detailed in the Fine-Tuning for Downstream Tasks subsection) and conducting an ablation study (detailed in the Ablations section) to validate the contribution of each architectural component.}
  \item Do you discuss what is ``the cost`` of misclassification and fault (in)tolerance?
    \answerYes{Yes. We discuss the cost of misclassification in our Case Studies, and in the Conclusions and Implications section.}
  
\end{enumerate}

\item Additionally, if you are using existing assets (e.g., code, data, models) or curating/releasing new assets, \textbf{without compromising anonymity}...
\begin{enumerate}
  \item If your work uses existing assets, did you cite the creators?
    \answerYes{Yes. We cite the creators of all datasets used (Dreaddit \cite{turcan-mckeown-2019-dreaddit}, Ruddit \cite{hendrickx2021ruddit}, NormVio \cite{park2021detecting}, and HatefulDiscussions \cite{hebert2023qualitative}) as well as software libraries PyTorch, PyTorch Geometric, and publicly available models such as EmbeddingGemma \cite{google2024embeddinggemma}.}
  \item Did you mention the license of the assets?
    \answerYes{Yes. While the specific datasets used do not have associated licenses, we acknowledge and adhere to the Reddit Terms of Service in our Ethical Statement.}
  \item Did you include any new assets in the supplemental material or as a URL?
    \answerYes{Yes. We include a link to our code in the Code section.}
  \item Did you discuss whether and how consent was obtained from people whose data you're using/curating?
    \answerNA{Yes. As stated in the Ethical Statement, all data used was publicly released prior to our collection, consistent with standard practices for public social media data analysis.}
  \item Did you discuss whether the data you are using/curating contains personally identifiable information or offensive content?
    \answerYes{Yes. We discuss the handling of personally identifiable information in our Ethical Statement.}
  \item If you are curating or releasing new datasets, did you discuss how you intend to make your datasets FAIR?
    \answerNA{We did not release a new dataset in this work.}
  \item If you are curating or releasing new datasets, did you create a Datasheet for the Dataset? 
    \answerNA{We did not release a new dataset in this work.}
\end{enumerate}

\item Additionally, if you used crowdsourcing or conducted research with human subjects, \textbf{without compromising anonymity}...
\begin{enumerate}
  \item Did you include the full text of instructions given to participants and screenshots?
    \answerNA{We did not crowdsource or conduct research with human subjects in this work.}
  \item Did you describe any potential participant risks, with mentions of Institutional Review Board (IRB) approvals?
    \answerNA{We did not crowdsource or conduct research with human subjects in this work.}
  \item Did you include the estimated hourly wage paid to participants and the total amount spent on participant compensation?
    \answerNA{We did not crowdsource or conduct research with human subjects in this work.}
  \item Did you discuss how data is stored, shared, and deidentified?
     \answerNA{We did not crowdsource or conduct research with human subjects in this work.}
\end{enumerate}

\end{enumerate}

\end{document}